\documentclass{SciPost}

\hypersetup{
    colorlinks,
    linkcolor={red!50!black},
    citecolor={blue!50!black},
    urlcolor={blue!80!black}
}

\usepackage[bitstream-charter]{mathdesign}
\usepackage[inline]{enumitem}
\usepackage{subcaption}
\usepackage[british]{babel}

\DeclareSymbolFont{usualmathcal}{OMS}{cmsy}{m}{n}
\DeclareSymbolFontAlphabet{\mathcal}{usualmathcal}

\fancypagestyle{SPstyle}{
\fancyhf{}
\lhead{\colorbox{scipostdeepblue}{\bf \color{white} ~SciPost Physics Proceedings }}
\rhead{{\bf \color{scipostdeepblue} ~Submission }}

\fancyfoot[C]{\textbf{\thepage}}
}

\begin{document}

\pagestyle{SPstyle}

\begin{center}{\Large \textbf{\color{scipostdeepblue}{
Event Tokenization and Masked-Token Prediction for Anomaly Detection at the Large Hadron Collider\\
}}}\end{center}

\begin{center}\textbf{
Ambre Visive\textsuperscript{1,2$\star$},
Polina Moskvitina \textsuperscript{3,2},
Clara Nellist \textsuperscript{1,2},
Roberto Ruiz de Austri\textsuperscript{4} and
Sascha Caron\textsuperscript{3,2}
}\end{center}

\begin{center}
{\bf 1} Institute of Physics, University of Amsterdam, Amsterdam,The Netherlands
\\
{\bf 2} Nikhef, Dutch National Institute for Subatomic Physics, Amsterdam, The Netherlands
\\
{\bf 3} High Energy Physics, Radboud University, Nijmegen, The Netherlands
\\
{\bf 4} Instituto de Física Corpuscular, IFIC-UV/CSIC, Paterna, Spain
\\[\baselineskip]
$\star$ \href{mailto:email1}{\small ambre.visive@cern.ch}\,
\end{center}

\definecolor{palegray}{gray}{0.95}
\begin{center}
\colorbox{palegray}{
  \begin{tabular}{rr}
  \begin{minipage}{0.37\textwidth}
    \includegraphics[width=60mm]{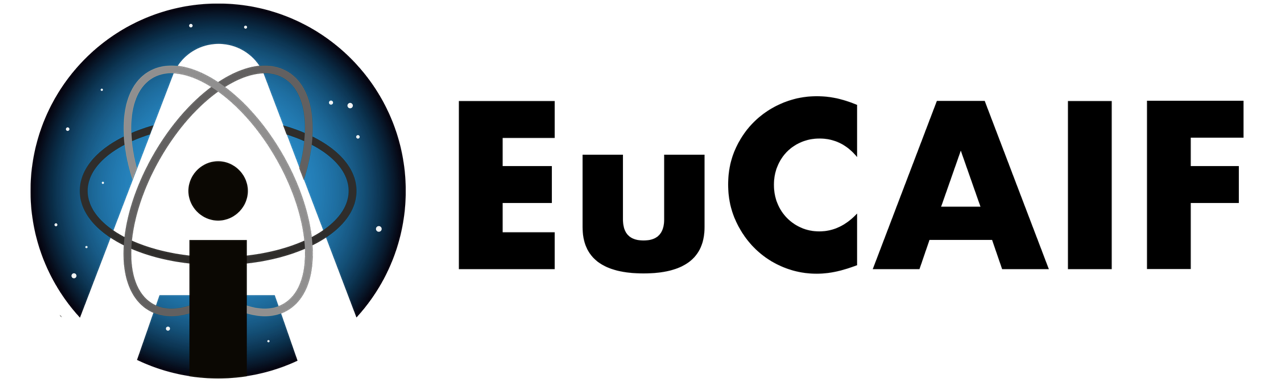}
  \end{minipage}
  &
  \begin{minipage}{0.5\textwidth}
    \vspace{5pt}
    \vspace{0.5\baselineskip} 
    \begin{center} \hspace{5pt}
    {\it The 2nd European AI for Fundamental \\Physics Conference (EuCAIFCon2025)} \\
    {\it Cagliari, Sardinia, 16-20 June 2025
    }
    \vspace{0.5\baselineskip} 
    \vspace{5pt}
    \end{center}
    
  \end{minipage}
\end{tabular}
}
\end{center}

\section*{\color{scipostdeepblue}{Abstract}}
\textbf{\boldmath{
We propose a novel use of Large Language Models (LLMs) as unsupervised anomaly detectors in particle physics. Using lightweight LLM-like networks with encoder-based architectures trained to reconstruct background events via masked-token prediction, our method identifies anomalies through deviations in reconstruction performance, without prior knowledge of signal characteristics. Applied to searches for simultaneous four-top-quark production, this token-based approach shows competitive performance against established unsupervised methods and effectively captures subtle discrepancies in collider data, suggesting a promising direction for model-independent searches for new physics.
}}

\vspace{\baselineskip}

\noindent\textcolor{white!90!black}{%
\fbox{\parbox{0.975\linewidth}{%
\textcolor{white!40!black}{\begin{tabular}{lr}%
  \begin{minipage}{0.6\textwidth}%
    {\small Copyright attribution to authors. \newline
    This work is a submission to SciPost Phys. Proc. \newline
    License information to appear upon publication. \newline
    Publication information to appear upon publication.}
  \end{minipage} & \begin{minipage}{0.4\textwidth}
    {\small Received Date \newline Accepted Date \newline Published Date}%
  \end{minipage}
\end{tabular}}
}}
}




\section{Introduction}
\label{sec:intro}
Large Language Models (LLMs), originally developed for natural language tasks \cite{OpenAi2017llm}, have shown remarkable capabilities in modeling complex data distributions \cite{{minaee2025largelanguagemodelssurvey},{lee2023lanobertloganomalydetection},{yang2025researchcloudplatformnetwork},{Pospieszny_2025}}. Their success is largely driven by transformer architectures \cite{DBLP:journals/corr/VaswaniSPUJGKP17} and large-scale training on diverse datasets \cite{stochasticparrotsBender2021}. In high-energy physics, the increasing data volume from the LHC opens new opportunities for applying such models \cite{atlas2025highlightshllhcphysicsprojections}. In this work, we explore LLM-like networks as potential unsupervised anomaly detection techniques, trained to reconstruct background events and identify deviations in the reconstruction without prior signal knowledge. This approach aims to improve sensitivity to rare Standard Model processes and uncover potential Beyond the Standard Model signatures.

\section{Dataset and Model}
\label{sec:DatasetModel}
\vspace{-0.25cm}
\subsection{Physics Motivation}
\label{sec:PhysicsMotivation}
Four-top-quark events present a complex final state, with 0-4 leptons and 4-12 jets, including four from bottom quarks, due to the dominant decay channel $t \rightarrow W + b$, as $|V_{tb}|^2 \sim 1$. Their signature closely resembles that of $t\bar{t}WW$ and $t\bar{t}W$, differing mainly by additional $b$-jets. Similarly, $t\bar{t}Z$, and $t\bar{t}H$ processes can be confused with four-top-quark signatures, as $Z$ and Higgs bosons often decay into jets or leptons. Due to these overlaps, we treat $t\bar{t}WW$, $t\bar{t}W$, $t\bar{t}Z$ and $t\bar{t}H$ as \textbf{background events} in this work, while $t\bar{t}t\bar{t}$ is the \textbf{signal process}, we are trying to isolate. Its rarity and complex topology make it an ideal benchmark for evaluating unsupervised anomaly detection methods on Standard Model (SM) processes.
\vspace{-0.25cm}
\subsection{The Dark Machines Datasets and Data Format} 
\label{sec: DMdata}
This study uses SM datasets from the Dark Machines challenge \cite{Aarrestad_2022},  comprising over $10^{9}$ simulated $pp$ collisions at $\sqrt{s} = 13~\mathrm{TeV}$. Events are generated with {\fontfamily{cmtt}\selectfont MG5\_aMC@NLO}~2.7, hadronized using {\fontfamily{cmtt}\selectfont Pythia}~8.239, and processed through the {\fontfamily{cmtt}\selectfont Delphes}~3.4.2 with modified ATLAS-Run 2 settings. The generated Monte-Carlo data, stored in ROOT format, were converted to CSV files following the event selection in \cite{Moskvitina:2025epjc}. In a second pre-processing step, the event-type is extracted and mapped to an integer as in Table~\ref{table:SMlabels} to create labeled datasets. Each event is represented as a sequence of particle-object tags (see Table~\ref{table:objectype}) and their four-momenta (energy $E$, transverse momentum $p_T$, pseudo-rapidity $\eta$, azimuthal angle $\phi$), ordered by type and $p_T$, and the missing transverse energy of the event ($\|E^{miss}_T\|$) and its azimuthal angle ($\phi_{E^{miss}_T}$). Bottom-quark jets are tagged without uncertainty. Sequences are padded to a fixed length of 18 particles-objects. The dataset is then split into training (80\%), validation (10\%), and test (10\%) subsets. Further details are available in \cite{Builtjes:2022usj, Moskvitina:2025epjc, Aarrestad_2022}.
\[ obj_1;  obj_2; ...; obj_{18}; \|E^{miss}_T\|; \phi_{E^{miss}_T}; E_1; p_{T,1}, \eta_1; \phi_1; E_2; p_{T,2}, \eta_2; \phi_2;...;  E_{18}; p_{T,18}, \eta_{18}; \phi_{18} \]

As the input to the model consists of batches of token sequences, each representing a particle physics event, the missing transverse energy, its azimuthal angle, and the particles, the latter being characterised by its type, charge, transverse momentum, pseudo-rapidity, and azimuthal angle, have to be encoded. Constructing these sequences from the dataset requires a dedicated tokenization step that is described in more details in Section \ref{sec: LLM}.
\begin{table}[h]
    \centering
    \begin{tabular}{ c|c|c|c|c|c } 
        SM process & $t\bar{t}t\bar{t}$ & $t\bar{t}H$ & $t\bar{t}W$ & $t\bar{t}WW$ & $t\bar{t}Z$ \\ 
        \hline
        {\fontfamily{cmtt}\selectfont process ID } & 1 & 2 & 3 & 4 & 5\\ 
    \end{tabular}
    \caption{Labels of SM processes.}
    \label{table:SMlabels}
\end{table}
\begin{table}[h]
\centering
    \begin{tabular}{ c|c|c|c|c|c|c|c } 
    Object & jet & b-tagged jet & positron & electron & muon & anti-muon & photon \\
    \hline
    symbol ID & {\fontfamily{cmtt}\selectfont j} &{\fontfamily{cmtt}\selectfont b} & {\fontfamily{cmtt}\selectfont e+} & {\fontfamily{cmtt}\selectfont e- } & {\fontfamily{cmtt}\selectfont mu+ } & {\fontfamily{cmtt}\selectfont mu-} & {\fontfamily{cmtt}\selectfont g } \\
    \hline
    {\fontfamily{cmtt}\selectfont tag } & 1 & 2 & 3 & 4 & 5 & 6 & 7\\ 
    \end{tabular}
    \caption{Tags of particle-objects.}
\label{table:objectype} 
\end{table}

\vspace{-0.5cm}
\subsection{Large-Language-Model for Anomaly Detection and Tokenization Strategy} 
\label{sec: LLM}
\begin{figure}[b]
    \centering
    \includegraphics[width=0.5\linewidth]{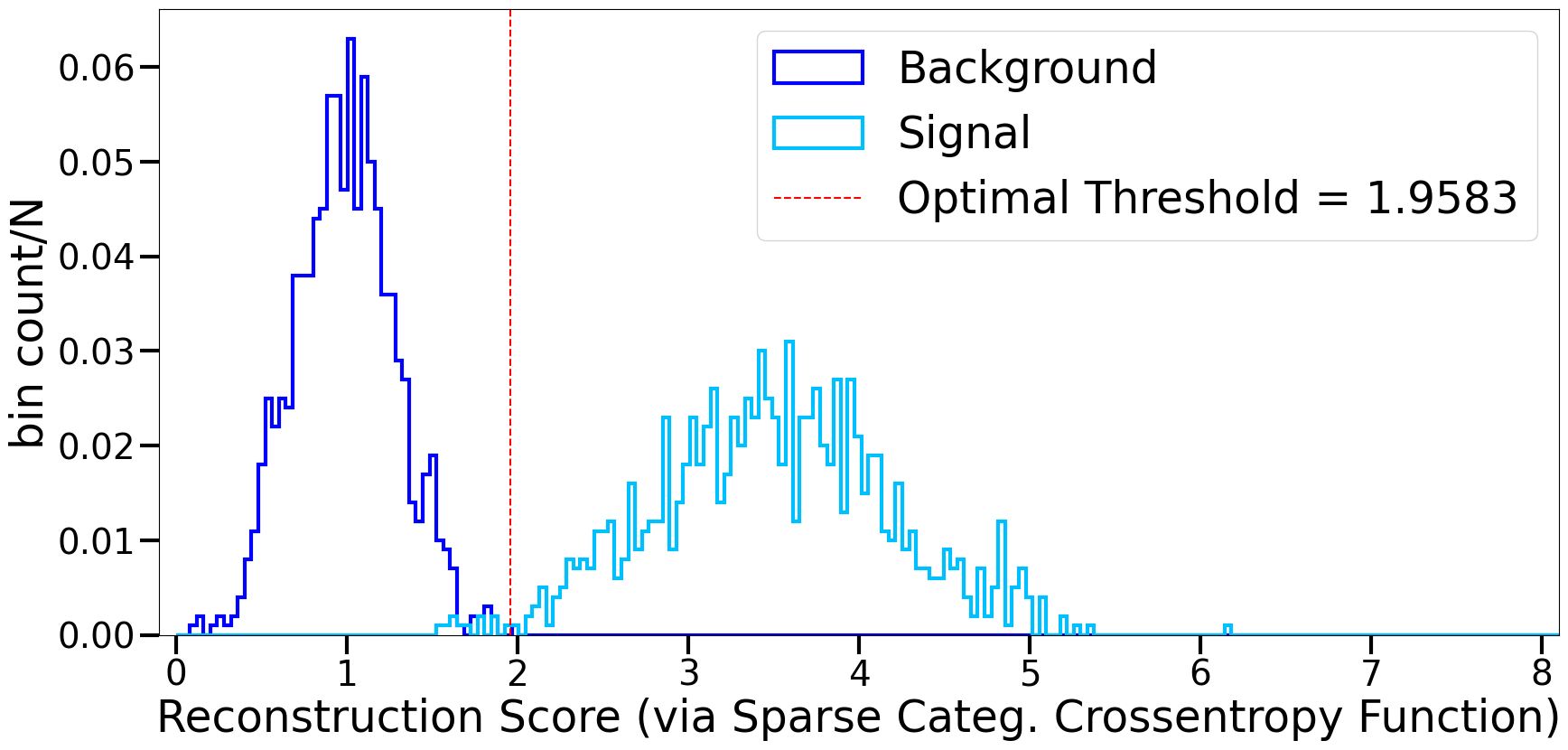}
    \caption{Illustrative distribution of the aggregated reconstruction scores in a perfectly trained model, evaluated with sparse categorical cross-entropy, for background (blue) and signal (cyan) events. The red dashed line indicates the optimal threshold used to separate the two classes.}
    \label{fig:perfectHisto}
\end{figure}

We employ a lightweight LLM-like model based on the transformer encoder. Input sequences, tokenized representations of particle physics events, are embedded. The core of the model consists of two transformer layers with four self-attention heads each, enabling the model to capture contextual relationships across tokens. It is followed by a linear projection and softmax layer that outputs a probability distribution over token classes, indicating the likelihood of each token being the correct reconstruction.
To have the model learn the distribution of background events, training is performed on background events only, using masked-token prediction, as introduced by BERT \cite{devlin2019bertpretrainingdeepbidirectional}: one token per event is randomly masked, and the model is trained to reconstruct it using Sparse Categorical Cross-Entropy loss. Optimization is done with \texttt{Adam}~\cite{kingma2017adam}, and early stopping is applied.
During inference, all tokens in each event are masked and reconstructed one at a time. The reconstruction scores are averaged to produce a reconstruction score per event. Events with poor reconstruction are flagged as anomalous, indicating deviation from the learned background distribution. This yields score distributions (see Figure~\ref{fig:perfectHisto}), from which thresholds can be defined to identify potential signal events.

A tokenization step is needed to represent particle physics events as sequences of {\fontfamily{cmtt}\selectfont tokens} suitable for LLM or LLM-like models. The tokens are obtained through a binning strategy: while particles are mapped to seven predefined categories (see Table~\ref{table:objectype}), different discretisation intervals for $p_T$, $\eta$, and $\phi$ were tested as careful selection of bin edges ensures meaningful separation of physical features. 
Performances of each binning-based tokenization strategy and the impact of including $\|E^{miss}_T\|$ and $\phi_{E^{miss}_T}$ as additional tokens in the sequence representation were assessed using a simple compact classifier neural network for rapid evaluation or the downstream model performance. In the most effective binning configuration:
\begin{itemize}
    \item[-]$p_T$, $\eta$ and $\|E^{miss}_T\|$ were each divided into 4 bins, with edges defined as containing 25\% of the background data in each bin.
    \item[-]$\phi$ and $\phi_{E^{miss}_T}$ were divided into 4 bins of width $\frac14\pi$.
    \item[-]With bins indexed from 1 onward, each particle token is defined as followed: \\
    \(token_{part}=64\times(bin_{obj}-1)+16\times(bin_{p_T}-1)+4\times(bin_{\eta}-1)+bin_{\phi}\).
    \item[-]It yields:
    \begin{itemize*}[itemjoin=\quad, before=\null, after=\hskip1.9em\hfill]
        \item[$\cdot$]\(token_{part}\in[1,448]\);
        \item[$\cdot$]\(token_{\|E^{miss}_T\|}\in[449,452]\);
        \item[$\cdot$]\(token_{\phi_{E^{miss}_T}}\in[453,456]\).
    \end{itemize*}
    \item[-]An event is represented as a sequence:\\
\([token_{part,1}, token_{part,2}, token_{part,3}, ..., token_{part,18}, token_{\|E^{miss}_T\|} , token_{\phi_{E^{miss}_T}}]\)
\end{itemize}
\noindent 
Since the original dataset is zero-padded and transformer models require uniform sequence lengths across batches, $token_{part,i}=0$ was set for padded entries where no particle is present.

This tokenization enables the model to effectively learn the structure of background events. However, it is possible that a deep-learned tokenization would yield better results.

\section{Results and Evaluation}
\label{sec: Results}
\vspace{-0.25cm}
\subsection{Search for Four-Top Production}
\label{sec: results4tops}

To evaluate the performance of the method on the simultaneous four-top-quarks production, both the background and the signal are tokenized using the strategy described in Subsection \ref{sec: LLM}. Once the model has been trained on the background data, both signal and background data are injected. Following the technique described in \ref{sec: LLM}, the distribution of the average reconstruction score of the signal and background events is obtained, as shown in Figure \ref{fig:4topHisto}. The common area of 70.85\% illustrates the overlap between the distributions, highlighting the model's ability to discriminate between background and signal events. The optimal threshold used to separate the two classes can be yielded and used as reference for future analysis. 
From this plot, the Receiving Operator Characteristic (ROC) curve can be derived (see Figure \ref{fig:ROC-AUC}) and the area under the curve (ROC-AUC) of the model can be calculated. The model yields: $\text{ROC-AUC}=0.67$.

\vspace{-0.25cm}
\subsection{Comparison with Established Unsupervised Methods}
\label{sec: comparison}

\begin{figure} [h!]
     \centering
     \begin{subfigure}[b]{0.48\textwidth}
         \centering
\includegraphics[width=\textwidth]{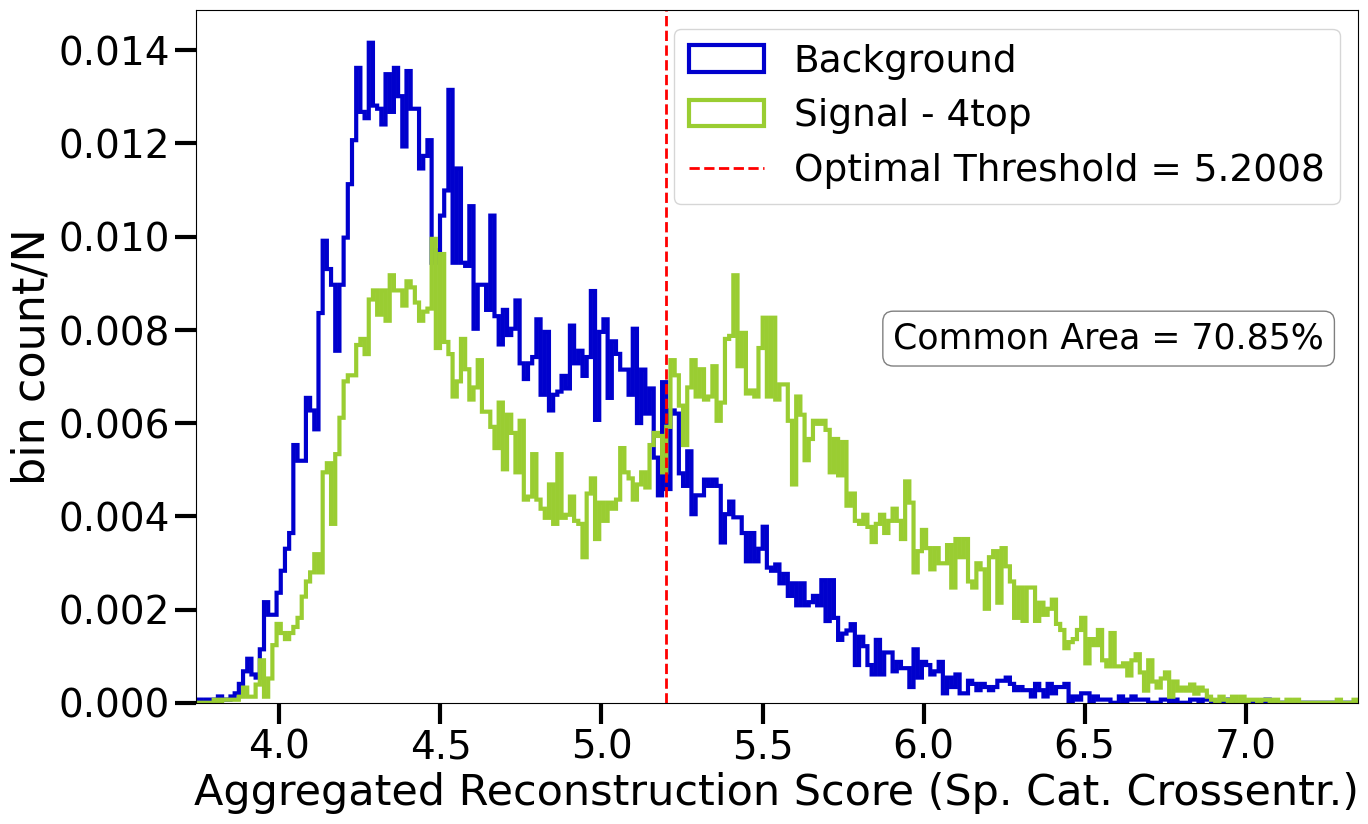}
         \caption{Distribution of the aggregated reconstruction scores, evaluated with a sparse categorical cross-entropy function, for background (blue) and four-top-signal (green) events. The red dashed line indicates the optimal threshold that can be used to best separate the two classes.}
         \label{fig:4topHisto}
     \end{subfigure}
     \hfill
     \begin{subfigure}[b]{0.48\textwidth}
         \centering
         \includegraphics[width=0.9\textwidth]{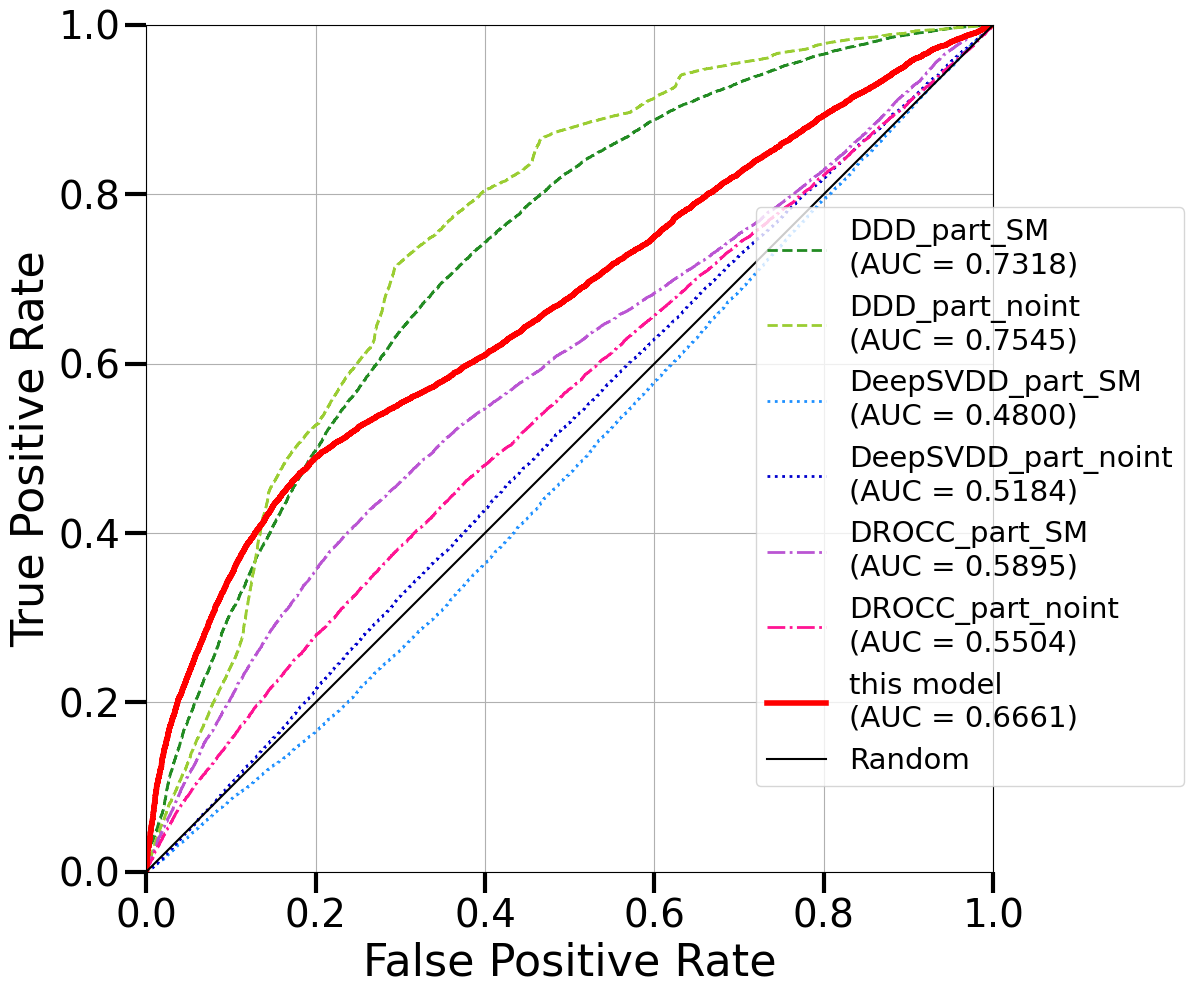}
         \caption{Comparison of the ROC curves of different unsupervised anomaly detection methods: the model presented in this paper (red), models presented in \cite{Moskvitina:2025epjc} (other colours).}
         \label{fig:ROC-AUC}
     \end{subfigure}
\caption{Results from the model and comparisons to alternative methods.}
\end{figure}

A comparison was conducted with established unsupervised anomaly detection methods implemented in \cite{Moskvitina:2025epjc}: the DDD, DeepSVDD and DROCC methods. As illustrated in Figure~\ref{fig:ROC-AUC}, although the proposed method does not surpass the DDD-based techniques, improved performance was observed relative to DeepSVDD and DROCC, indicating competitive behavior in unsupervised techniques for anomaly detection.

\vspace{-1cm}
  
\section{Conclusion}
\label{sec: Conclusion}

A novel application of LLM-like models for unsupervised anomaly detection in particle physics has been presented. Although the results remain preliminary, promising performance was observed in identifying rare processes such as four-top-quark production. While not outperforming all existing approaches, competitive results were achieved, supported by a flexible token-based representation of collider data. With further optimization of the tokenization scheme and model architecture, improvements in sensitivity and robustness are anticipated, making this approach a viable candidate for model-independent searches in future high-energy physics analyses.



\bibliography{SciPost_BiBTeX_File.bib}


\end{document}